\documentclass[
    11pt,
	amsmath,amssymb,
	superscriptaddress,
	onecolumn
   ]{quantumarticle}

\newcommand{\expect}[1]{
    \ensuremath{\langle{#1}\rangle}
}

\usepackage[numbers,sort&compress]{natbib}
\usepackage{graphicx}
\usepackage{color}
\usepackage{dcolumn}
\usepackage{siunitx}
\usepackage{bm}
\usepackage{braket}
\begin{document}


\title[]{Theory of analytical energy derivatives for the variational quantum eigensolver}

\author{Kosuke Mitarai}
\email{u801032f@ecs.osaka-u.ac.jp}
\affiliation{Graduate School of Engineering Science, Osaka University, 1-3 Machikaneyama, Toyonaka, Osaka 560-8531, Japan.}
\affiliation{QunaSys Inc., High-tech Hongo Building 1F, 5-25-18 Hongo, Bunkyo, Tokyo 113-0033, Japan}

\author{Yuya O. Nakagawa}
\affiliation{QunaSys Inc., High-tech Hongo Building 1F, 5-25-18 Hongo, Bunkyo, Tokyo 113-0033, Japan}

\author{Wataru Mizukami}
\affiliation{Interdisciplinary Graduate School of Engineering Science, Kyushu University, 6-1 Kasuga-Koen, Kasuga, Fukuoka, 816-6580, Japan.}

\date{\today}

\begin{abstract}
    The variational quantum eigensolver (VQE) and its variants, which is a method for finding eigenstates and eigenenergies of a given Hamiltonian, are appealing applications of near-term quantum computers.
Although the eigenenergies are certainly important quantities which determine properties of a given system, their derivatives with respect to parameters of the system, such as positions of nuclei if we target a quantum chemistry problem, are also crucial to analyze the system. 
Here, we describe methods to evaluate analytical derivatives of the eigenenergy of a given Hamiltonian, including the excited state energy as well as the ground state energy, with respect to the system parameters in the framework of the VQE.
We give explicit, low-depth quantum circuits which can measure essential quantities to evaluate energy derivatives, incorporating with proof-of-principle numerical simulations.
This work extends the theory of the variational quantum eigensolver, by enabling it to measure more physical properties of a quantum system than before and to explore chemical reactions.
\end{abstract}

\maketitle

\section{Introduction}

The variational quantum eigensolver (VQE) has attracted much attention as a potential application of near-term quantum computers \cite{Peruzzo2013}.
The VQE is an iterative algorithm to construct a quantum circuit that outputs eigenstates and eigenenergy of a Hamiltonian which describes the system under consideration. 
Originally, the method was devised for finding the ground state of a system \cite{Peruzzo2013}. 
It has subsequently been extended for excited states by several proposals \cite{Nakanishi2018, Endo2018b, Higgott2018, McClean2017b}.
From the generated eigenstates, one can measure its associated physical quantities, such as the particle densities and transition amplitudes between the different eigenstates.

Though eigenenergy and associated particle density are certainly important quantities, the wavefunction of a quantum system has valuable information besides those.
Quantum chemistry calculations, which would be one of the most promising applications of the VQE, often utilize such information.
Among such, we focus on energy derivatives in this work.
Many time-independent physical/chemical properties can be defined as derivatives of the energy \cite{Jensen2017,Gauss2004,Helgaker2012,Pulay2014}.
For example, the first derivatives of the energy with respect to nuclear coordinates give us the forces acting on atoms, which can be utilized for the task of locating energy extrema on the potential energy surface  (i.e., geometry optimization) \cite{Schlegel2011}.
The second-order derivatives give the force-constant matrix that not only helps to locate and verify transition states but also allows us to compute vibrational frequencies and partition functions within the harmonic approximation \cite{Schlegel2011,Wilson1980}. 
The energy derivatives with respect to external fields have to do with various spectroscopy: 
Intensities of infrared and Raman spectroscopy are proportional to the cross derivatives with respect to vibrational normal modes and external electric fields \cite{Wilson1980}; 
NMR chemical shifts can be obtained using the cross derivatives with respect to nuclear spin and magnetic fields \cite{Ramsey1950,Helgaker1999}.
Computing such derivatives is a core part of simulations or analysis of molecular spectra. 
If quantum chemical methods were unable to extract those properties, presumably they would not be widely used as it is today.

A simple way to compute energy derivatives is to use the finite difference method and calculate them numerically. 
This approach, however, suffers from high computational costs as well as numerical errors and instabilities \cite{Gauss2004,Pulay2014}. 
Say, the number of energy points needed to evaluate the forces increases linearly to the number of atoms. 
This high computational cost makes the numerical approach impractical in many cases.
Moreover, with near-term quantum devices, where noise is inevitable, the numerical difference approach would give us poor results.
The other way -- analytical approach -- is therefore vital. 
The theory and program codes of analytical energy derivatives indeed support the high practicality (and popularity) of today's molecular electronic structure theory \cite{Pulay2014}.
Methods to calculate the derivatives of excited state energy on classical computers has also been widely developed \cite{Foresman1992,Maurice1999,VanCaillie1999,VanCaillie2000,Furche2002,Bernard2012,Kohn2003,Hattig2005,Tew2019,Stanton1993,Stanton1994,Stanton1995,Nakajima1999,Kallay2004,Nakano1998, Celani2003, Yamaguchi1994, MacLeod2015,Sand2017,Hohenstein2015,Fales2017}, but still suffers from its high computational cost and relatively low accuracy.
The task of computing such derivatives on quantum computers has been addressed in the traditional methods which utilize the quantum phase estimation \cite{Kassal2009}, but not for the VQE.

In this work, we derive the analytical formulae and explicit quantum circuits to address the task of measuring the energy derivatives.
More specifically, we describe the methods to obtain the derivatives of the energy with respect to the system parameters up to the third order, from which one can extract the physical properties.
We also present a method to extract the derivatives of excited state energy based on the technique presented in Refs. \cite{Endo2018b,Higgott2018}.
Proof-of-principle numerical simulations are also performed to verify the correctness of the derived equation and circuits.
The presented methods extend the applicability of the VQE by enabling it to evaluate more physical properties than before.

\section{Variational quantum eigensolver}
Here, we briefly review the algorithm of the VQE.
In the VQE, we construct a parameterized quantum circuit $U(\theta)$ and the corresponding ansatz state $\ket{\psi(\theta)} = U(\theta)\ket{0}^{\otimes n}$, where $\ket{0}^{\otimes n}$ is an initialized $n$-qubit state and $\theta = (\theta_1,\ldots,\theta_{N_\theta}) \in \mathbb{R}^{N_\theta}$ is a vector of parameters implemented on the circuit with $N_\theta$ being the number of them.
A set of parameters $\theta$ are variationally optimized so that the expectation value $E(\theta) = \bra{\psi(\theta)}H\ket{\psi(\theta)}$ of a given Hamiltonian $H$ is minimized.
At the optimal point, one naturally expects
\begin{equation}\label{eq:grad_vanish}
    \frac{\partial E(\theta)}{\partial \theta_a} = 0,
\end{equation}
for all $a$.
We denote such an optimal point by $\theta^*$.
Let
\begin{equation}
    \ket{\partial_a \psi(\theta)} = \frac{\partial}{\partial \theta_a}\ket{\psi(\theta)}.
\end{equation}
The condition of Eq. (\ref{eq:grad_vanish}) reduces to
\begin{equation}
    \mathrm{Re}\bra{\psi(\theta^*)}H\ket{\partial_a \psi(\theta^*)} = 0.
\end{equation}
Higher derivatives of the wave function will be denoted by
\begin{equation}
    \ket{\partial_a\partial_b \cdots \partial_c \psi(\theta)} = \frac{\partial}{\partial \theta_a}\frac{\partial}{\partial \theta_b}\cdots\frac{\partial}{\partial \theta_c}\ket{\psi(\theta)},
\end{equation}
to shorten the notation.

As stated in the introduction, many physical properties of a quantum system are calculated from the derivative of the energy with respect to the system parameter in a given Hamiltonian.
The parameter can be, for example, the coordinates of atoms or the electric field applied to the system.
We denote such parameters by an $N_x$-dimensional vector $x\in \mathbb{R}^{N_x}$.
In this case, both of the Hamiltonian $H$ and the optimal parameter $\theta^*$ of the wave function at the specific value of $x$ is also a function of $x$, which will be denoted by $H(x)$ and $\theta^*(x)$, respectively.
Corresponding to the change of this problem setting, we redefine the energy as,
\begin{align}
    E(\theta, x) = \braket{\psi(\theta)|H(x)|\psi(\theta)}.
\end{align}

Let the optimal ground state energy be $E^*(x)$, that is, $E^*(x) = E(\theta^*(x),x)$.
In the following sections, we show the analytical forms of the derivatives such as $\frac{\partial E^*}{\partial x_i}$ and $\frac{\partial^2 E^*}{\partial x_i\partial x_j}$, which are the essential quantities for extracting physical properties of the target system.

\section{Analytical expression of derivatives}
The derivation of the formulae presented in this section is in the Appendix for completeness, or you can also refer to Ref.  \cite{Pulay2007}.
In the following, we use the following notation,
\begin{equation}
    \frac{\partial}{\partial \theta_a}\frac{\partial E(\theta^*(x),x)}{\partial x_j} :=\left. \frac{\partial}{\partial \theta_a}\frac{\partial E(\theta,x)}{\partial x_j}\right|_{\theta=\theta^*(x), x=x}.
\end{equation}
and likewise for terms similar to this.

\subsection{Derivatives of ground state energy}
The analytical expressions for the derivatives of ground state energy are the following,
\begin{align}
    \frac{\partial E^*(x)}{\partial x_i} &= \braket{\psi\left(\theta^*(x)\right)|\frac{\partial H(x)}{\partial x_i}| \psi\left(\theta^*(x)\right)}, \label{eq:energy_first_derivative}\\
    \frac{\partial}{\partial x_i}\frac{\partial E^*(x)}{ \partial x_j} &= \sum_a \frac{\partial \theta^*_a(x)}{\partial x_i}\frac{\partial}{\partial \theta_a}\frac{\partial E(\theta^*(x),x)}{\partial x_j} + \bra{\psi(\theta^*(x))} \frac{\partial }{\partial x_i}\frac{\partial H (x)}{\partial x_j} \ket{\psi(\theta^*(x))},\label{eq:energy_second_derivative}\\
    \frac{\partial}{\partial x_i}\frac{\partial}{\partial x_j}\frac{\partial E^*(x)}{ \partial x_k}&=
    \sum_{a,b,c}\frac{\partial}{\partial \theta_a}\frac{\partial}{\partial \theta_b}\frac{\partial E (\theta^*(x),x)}{\partial \theta_c}\frac{\partial \theta_a^*(x)}{\partial x_i}\frac{\partial \theta_b^*(x)}{\partial x_j}\frac{\partial \theta^*_c(x)}{\partial x_k} \nonumber \\
    &\quad + \bra{\psi(\theta^*(x))}\frac{\partial }{\partial x_i}\frac{\partial }{\partial x_j}\frac{\partial H (x)}{\partial x_k} \ket{\psi(\theta^*(x))} \nonumber \\
    &\quad + \sum_{a,b}\left[ \frac{\partial \theta^*_a(x)}{\partial x_i} \frac{\partial\theta^*_b(x)}{\partial x_j}\frac{\partial}{\partial \theta_b}\frac{\partial}{\partial \theta_a}\frac{\partial E (\theta^*(x),x)}{\partial x_k}
    \right.\nonumber \\
    &\qquad\qquad +\frac{\partial \theta^*_a(x)}{\partial x_k} \frac{\partial\theta^*_b(x)}{\partial x_i}\frac{\partial}{\partial \theta_b}\frac{\partial}{\partial \theta_a}\frac{\partial E(\theta^*(x),x)}{\partial x_j}\nonumber\\
    &\left.\qquad\qquad +\frac{\partial \theta^*_a(x)}{\partial x_j} \frac{\partial\theta^*_b(x)}{\partial x_k}\frac{\partial}{\partial \theta_b}\frac{\partial}{\partial \theta_a}\frac{\partial E(\theta^*(x),x)}{\partial x_i}
    \right]\nonumber\\
    &\quad + \sum_a \left[\frac{\partial \theta^*_a(x)}{\partial x_i} \frac{\partial}{\partial \theta_a}\frac{\partial }{\partial x_j}\frac{\partial E(\theta^*(x),x)}{\partial x_k} \right.\nonumber \\
    &\quad\quad\quad\quad + \frac{\partial \theta^*_a(x)}{\partial x_k} \frac{\partial}{\partial \theta_a}\frac{\partial }{\partial x_i}\frac{\partial E(\theta^*(x),x)}{\partial x_j}\nonumber\\
    &\quad\quad\quad\quad \left.+ \frac{\partial \theta^*_a(x)}{\partial x_j}\frac{\partial}{\partial \theta_a}\frac{\partial }{\partial x_k}\frac{\partial E(\theta^*(x),x)}{\partial x_i}\right],\label{eq:energy_third_derivative}
\end{align}
where we assumed $\frac{\partial E(\theta^*(x),x)}{\partial \theta}=0$.
Note that, in general, the formulae for the $d$-th derivative of $E^*(x)$ contains $\theta$-detrivatives up to the $d$-th in the form of $\frac{\partial^q }{\partial \theta^q}\frac{\partial^{d-q} E}{\partial x^{d-q}}$ for $q=1,\ldots,d$.
Also, Wigner's $(2n+1)$-rule \cite{Pulay2007} ensures that $x$-derivatives of the optimal parameter, $\theta^*(x)$, up to the $n$-th are sufficient for calculating $(2n+1)$-th derivative of $E^*(x)$.
In other words, for $d$-th derivative of $E^*(x)$ we only need $\lfloor d/2 \rfloor$-th derivative of $\theta^*(x)$, where $\lfloor y \rfloor$ is the floor function denoting the greatest integer less than or equal to $y$.
The term $\frac{\partial \theta^*_a(x)}{\partial x_i}$ in the above equation can be calculated by solving the response equation, which we write down in the next subsection.

\subsection{Derivatives of optimal parameters}
The first and second derivatives of the optimal parameter, $\theta^*(x)$, can be obtained from the following response equation. 
\begin{align}
    \sum_b \frac{\partial}{\partial \theta_a}\frac{\partial E(\theta^*(x),x)}{\partial \theta_b}\frac{\partial \theta^*_b(x)}{\partial x_i} &= -\frac{\partial}{\partial \theta_a}\frac{\partial E(\theta^*(x),x)}{\partial x_i}, \label{eq:theta_first_derivative}\\
    \sum_b \frac{\partial}{\partial \theta_a}\frac{\partial E(\theta^*(x),x)}{\partial \theta_b}\frac{\partial}{\partial x_i}\frac{\partial \theta^*_b(x)}{\partial x_j} &= -\gamma^{(ij)}_a(\theta^*(x),x), \label{eq:theta_second_derivative}
\end{align}
where,
\begin{align}
    &\gamma_c^{(ij)} = \sum_{a,b}\frac{\partial }{\partial \theta_c}\frac{\partial }{\partial \theta_a}\frac{\partial E}{\partial \theta_b} \frac{\partial \theta^*_a}{\partial x_i}\frac{\partial \theta^*_b}{\partial x_j}
    + 2\sum_{a} \frac{\partial }{\partial \theta_c}\frac{\partial}{\partial \theta_a}\frac{\partial E}{\partial x_j}\frac{\partial \theta^*_a}{\partial x_i}
    +\frac{\partial}{\partial \theta _c}\frac{\partial}{\partial x_i}\frac{\partial E}{\partial x_j}.
\end{align}
By Wigner's $(2n+1)$-rule \cite{Pulay2007}, the above equations are enough to obtain derivatives of the energy up to the 5th-order.

\section{Measurement and calculation of derivatives of ground state}\label{sec:ground_state_derivative}
In this section, we describe the methodology for calculating the derivatives of the ground state energy, whose analytical expressions are shown in the previous section, when given an optimal circuit parameter $\theta^*(x)$ at some $x$ that gives the local minimum of $E(\theta,x)$.

\subsection{Notations and assumptions}\label{sec:notation_assumption}
In the VQE, we target a Hamiltonian which acts on an $n$-qubit system and is decomposed into a sum of Pauli strings, $\mathcal{P} = \{I,X,Y,Z\}^{\otimes n}$, as follows,
\begin{align}
    H(x) = \sum_{P\in\mathcal{P}} h_P(x)P, \label{eq:hamiltonian}
\end{align}
where $h_P(x)\in \mathbb{R}$. $h_P(x)$ is assumed to be non-zero only for $O(\mathrm{poly}(n))$ terms.
For a quantum chemistry problem, the original Hamiltonian has the following form,
\begin{align}
    H(x) = \sum_{i,j} h_{ij}(x)c_i^\dagger c_j + \sum_{i,j,k,l} h_{ijkl}(x)c_i^\dagger c^\dagger_j c_k c_l \label{eq:electron_hamiltonian},
\end{align}
where $c_i^\dagger$ and $c_i$ are the fermion creation and annihilation operators acting on the $i$-th orbital, respectively.
Eq. (\ref{eq:electron_hamiltonian}) is converted to the form of Eq. (\ref{eq:hamiltonian}) by, for example, Jordan-Wigner or Braviy-Kitaev transformation \cite{Seeley2012, McArdle2018a}.
Note that because we always work in the second quantization formalism, the effect of the change of the molecular orbital corresponding to the change of molecular geometry is totally absorbed in the coefficients $h(x)$.
Therefore, the change of the molecular orbital does not explicitly appear in the following discussion.

To calculate the energy derivatives of such Hamiltonian, first of all, we assume that the derivatives of Hamiltonian, $\frac{\partial H}{\partial x_i}$, $\frac{\partial }{\partial x_i}\frac{\partial H}{\partial x_j}$, and so on, can be calculated by the classical computer, e.g., the conventional libraries of quantum chemistry calculation.
In other words, we are able to calculate the quantities such as $\frac{\partial h_P(x)}{\partial x_i}$ and $\frac{\partial }{\partial x_i}\frac{\partial h_P(x)}{\partial x_j}$.
For quantum chemisry problems given in terms of Hartree-Fock orbitals, these calculations correspond to solving the coupled perturbed Hartree-Fock equation \cite{Pulay2007, Yamaguchi1994}.

Notice that under this assumption, we only need to consider how to evaluate quantities which involve the differentiation with respect to $\theta$ such as $\frac{\partial}{\partial \theta_a}\frac{\partial}{\partial \theta_b}\frac{\partial E}{\partial x_i} $, because the expectation values such as $\frac{\partial}{\partial x_i}\frac{\partial}{\partial x_j}\frac{\partial E}{\partial x_k} = \bra{\psi(\theta)}\frac{\partial }{\partial x_i}\frac{\partial }{\partial x_j}\frac{\partial H(x)}{\partial x_k} \ket{\psi(\theta)}$ at $\theta=\theta^*(x)$, can be evaluated with the exactly same procedure as the usual VQE, i.e., we can measure the expectation value of each Pauli term which appears in $\frac{\partial }{\partial x_i}\frac{\partial }{\partial x_j}\frac{\partial H(x)}{\partial x_k}$.
After the measurement of all quantities which appear in Eqs.~(\ref{eq:energy_first_derivative}), (\ref{eq:energy_second_derivative}), and (\ref{eq:energy_third_derivative}), one can compute the energy derivative using a classical computer by summing up the terms.

The parameterized quantum state, $\ket{\psi(\theta)}$, is constructed by applying a parameterized unitary matrix, that is, a parameterized quantum circuit, $U(\theta)$ to an initialized state; $\ket{\psi(\theta)} = U(\theta)\ket{0}^{\otimes n}$.
Following \cite{Li2016c}, we assume $U(\theta)$ to be a product of unitary matrices each with one parameter,
\begin{equation}
    U(\theta) = U_{N_\theta}(\theta_{N_\theta})\cdots U_2(\theta_2)U_1(\theta_1).
\end{equation}
We define each unitary $U_a(\theta_a)$ to be generated by a generator $G_a$ as $U_a=e^{i\theta_a G_a}$, which can be decomposed into a sum of Pauli strings;
\begin{equation}
    G_a = \sum_{\mu} g_{a,\mu}P_{a,\mu},
\end{equation}
where $g_{a,\mu}\in \mathbb{R}$ and $P_{a,\mu} \in \mathcal{P}$.
We often use this form of parametrized quantum circuits, for example see Ref.~\cite{Dallaire-Demers2018, Kandala2017, Nam2019, Grimley2018}.

\subsection{Overview of the algorithm}\label{sec:overall}
The presented algorithm in this work for evaluating the $d$-th derivative is the following:
\begin{enumerate}
    \item Perform the VQE and obtain the optimal parameter $\theta^*(x)$.
    \item Compute derivatives of the Hamiltonian, $\frac{\partial^q H(x)}{\partial x^q}$, for $q=1,\ldots, d$ on a classical computer.
    \item Evaluate $x$-derivatives of $E$, $\frac{\partial^q E}{\partial x^q}$, for $q=1,\ldots, d$ at $\theta=\theta^*(x)$ by the method described above (Sec \ref{sec:notation_assumption}).
    \item Evaluate terms involving differentiations with respect to $\theta$, $\frac{\partial^q }{\partial \theta^q}\frac{\partial^{d-q} E}{\partial x^{d-q}}$ and $\frac{\partial^q E}{\partial \theta^q}$, for $q=1,\ldots,d$ at $\theta=\theta^*(x)$ on a quantum device by the method described in the following subsections (Secs. \ref{sec:theta_second_derivative}-\ref{sec:theta_x_general_derivative}).
    \item Solve the response equations to obtain $\frac{\partial^q \theta^*}{\partial x^{q}}$ for $q=1,\ldots,\lfloor d/2\rfloor$. (For $d\leq3$, solving Eq.(\ref{eq:theta_first_derivative}) suffices.)
    \item Substitute all terms into Eqs.~(\ref{eq:energy_first_derivative}), (\ref{eq:energy_second_derivative}), and (\ref{eq:energy_third_derivative}) to obtain the energy derivatives.
\end{enumerate}

Let us recall $N_H$ and $N_{\theta}$ be the number of terms in the target Hamiltonian and the number of the VQE parameters.
The cost to evaluate $d$-th derivatives by the above algorithm is roughly $O(N_H N_\theta^{d})$ when we ignore the cost of the task performed on a classical computer.
This is because the most time-consuming part of the algorithm is Step 3 and 4 on a quantum computer, which take $O(N_H N_\theta^{d})$ as shall be clear in Secs. \ref{sec:theta_second_derivative}-\ref{sec:theta_x_general_derivative} and \ref{sec:cost}.

\subsection{Measurement of $\frac{\partial}{\partial \theta_a}\frac{\partial E}{\partial \theta_b}$} \label{sec:theta_second_derivative}
The key quantities for evaluating $\frac{\partial }{\partial x_i}\frac{\partial E^*}{\partial x_j}$ and higher order derivatives are $\frac{\partial}{\partial \theta_a}\frac{\partial}{\partial \theta_b} \cdots \frac{\partial E}{\partial \theta_c} $ and $\frac{\partial}{\partial \theta_a}\frac{\partial}{\partial \theta_b} \cdots \frac{\partial}{\partial \theta_c} \frac{\partial E}{\partial x_i}$.
Note that the first order derivative, $\frac{\partial E}{\partial \theta_a}$ can be evaluated by the method presented in Ref. \cite{Mitarai2018}.  
First, we show how to measure $\frac{\partial}{\partial \theta_a}\frac{\partial E}{\partial \theta_b}$.
A detailed expression of $\frac{\partial}{\partial \theta_a}\frac{\partial E}{\partial \theta_b}$ is, 
\begin{align}
    \frac{\partial}{\partial \theta_a}\frac{\partial E (\theta^*(x),x) }{\partial \theta_b} 
    &= 2\mathrm{Re}\left[\braket{\partial_a\partial_b\psi(\theta^*(x))|H(x)|\psi(\theta^*(x))}+\braket{\partial_a\psi(\theta^*(x))|H(x)|\partial_b\psi(\theta^*(x))}\right].
\end{align}
$\ket{\partial_a \psi(\theta)}$ can be expressed as,
\begin{align}
    \ket{\partial_a \psi(\theta)} 
    &= i\sum_{\mu} g_{a,\mu} U_N(\theta_N)\cdots P_{a,\mu}U_a(\theta_a)\cdots U_2(\theta_2)U_1(\theta_1)\ket{0}^{\otimes n},
\end{align}
and $\ket{\partial_a\partial_b \psi(\theta)}$ is,
\begin{align}
    \ket{\partial_a \partial_b \psi(\theta)} 
    &= -\sum_{\mu,\nu} g_{a,\mu}g_{b,\nu} U_N(\theta_N)\cdots P_{a,\mu}U_a(\theta_a)\cdots P_{b,\nu}U_b(\theta_b) \cdots U_2(\theta_2)U_1(\theta_1)\ket{0}^{\otimes n}.
\end{align}
For convinience, we define,
\begin{align}
    \ket{\phi_{(a,\mu),(b,\nu),\cdots(c,\rho)}(\theta)} := U_N(\theta_N)\cdots (iP_{a,\mu})U_a(\theta_a)\cdots (iP_{b,\nu})U_b(\theta_b) \cdots (iP_{c,\rho})U_c(\theta_c) \cdots U_1(\theta_1)\ket{0}^{\otimes n}. \label{eq:def_phi}
\end{align}
Then,
\begin{align}
    &\frac{\partial}{\partial \theta_a}\frac{\partial E (\theta^*(x),x)}{\partial \theta_b}  \nonumber\\ 
    &= 2\sum_{\mu,\nu}\sum_{Q\in\mathcal{P}} h_Q(x) g_{a,\mu}g_{b,\nu}\mathrm{Re}\left[\bra{\phi_{(a,\mu),(b,\nu)}(\theta^*(x))}Q\ket{\psi(\theta^*(x))}+
    \braket{\phi_{(a,\mu)}(\theta^*(x))|Q|\phi_{(b,\nu)}(\theta^*(x))}\right].\label{eq:nabla_theta_2_E}
\end{align}

\begin{figure}
    \centering
    \includegraphics[width=0.7\linewidth]{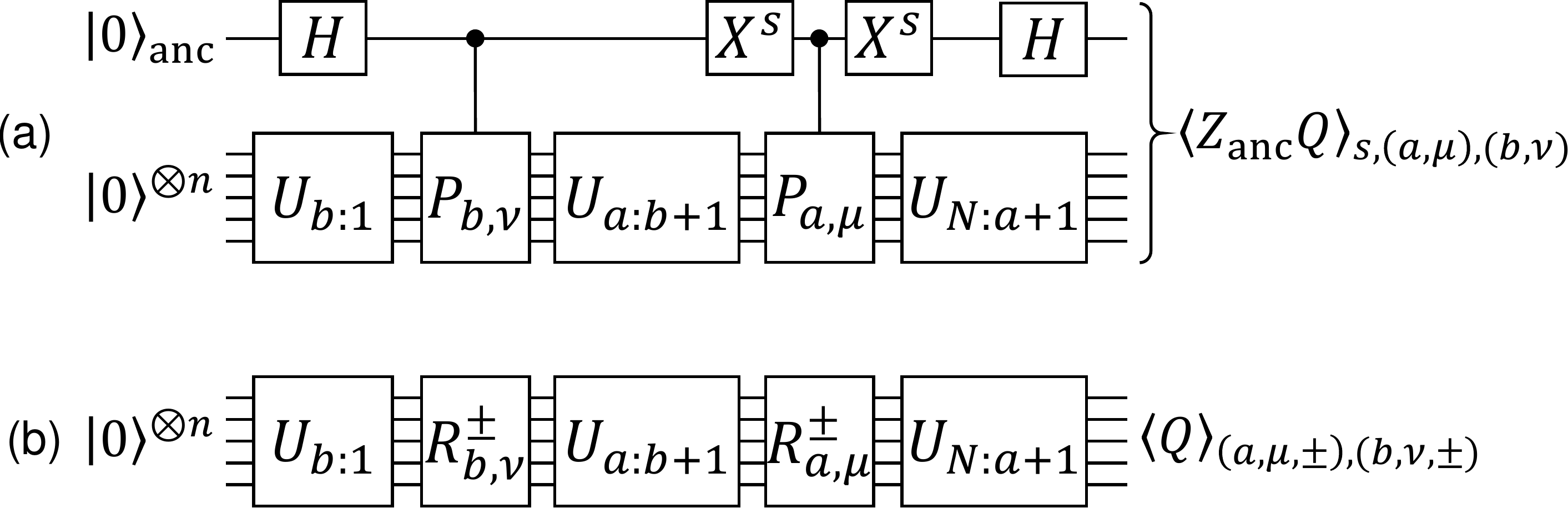}
    \caption{\label{fig:second_derivative} Quantum circuit to evaluate $\frac{\partial}{\partial \theta_a}\frac{\partial}{\partial \theta_b} E(\theta^*(x),x)$. $U_{a:b} = U_a\cdots U_{b+1}U_b$. (a) Ancilla-based approach. This circuit is a variant of the one presented in Ref.~\cite{Li2016c}. In the figure, $Z_{\mathrm{anc}}$ is the Pauli $Z$ operator acting only on the ancilla qubit, and $s\in\{0,1\}$. See Eq.~(\ref{eq:second_derivative_with_ancilla}) for detailed procedure. (b) Low-depth version of (a), derived with the strategy presented in Ref.~\cite{Mitarai2019a}. In the figure, $R_{a,\mu}^{\pm} = \exp(\pm i\pi P_{a,\mu}/4)$. See Eq.~(\ref{eq:second_derivative_low-depth}) for detailed procedure.
    }
\end{figure}

In Fig.~\ref{fig:second_derivative} (a), we show quantum circuits for evaluation of each of the terms in the above equation, which is a variant of the circuit presented in Ref.~\cite{Li2016c}. 
From the outputs of the circuits in Fig.~\ref{fig:second_derivative} (a), each term of Eq.~(\ref{eq:nabla_theta_2_E}) can be calculated by,
\begin{align}
    \mathrm{Re}[\braket{\phi_{(a,\mu),(b,\nu)}(\theta^*(x))|Q|\psi(\theta^*(x))}] &= \expect{Z_{\mathrm{anc}}Q}_{0,(a,\mu),(b,\nu)} \nonumber \\
    \mathrm{Re}[\braket{\phi_{(a,\mu)}(\theta^*(x))|Q|\phi_{(b,\nu)}(\theta^*(x))}] &= \expect{Z_{\mathrm{anc}}Q}_{1,(a,\mu),(b,\nu)} \label{eq:second_derivative_with_ancilla}
\end{align}

The strategy proposed in Ref.~\cite{Mitarai2019a} gives low-depth versions of the circuits to measure the same quantities.
We show the low-depth quantum circuit in Fig.~\ref{fig:second_derivative} (b).
From the measurement of $\expect{Q}_{(a,\mu,\pm),(b,\nu,\pm)}$ with the circuit in Fig.~\ref{fig:second_derivative} (b), we can evaluate each term of Eq.~(\ref{eq:nabla_theta_2_E}) by the following formula,
\begin{align}
    &2\mathrm{Re}\left[\bra{\phi_{(a,\mu),(b,\nu)}(\theta^*(x))}Q\ket{\psi(\theta^*(x))}+
    \braket{\phi_{(a,\mu)}(\theta^*(x))|Q|\phi_{(b,\nu)}(\theta^*(x))}\right]\nonumber \\
    &\quad = \expect{Q}_{(a,\mu,+),(b,\nu,+)} + \expect{Q}_{(a,\mu,-),(b,\nu,-)} - \expect{Q}_{(a,\mu,-),(b,\nu,+)} - \expect{Q}_{(a,\mu,+),(b,\nu,-)}.
    \label{eq:second_derivative_low-depth}
\end{align}

Notice that the low-depth version doubles the number of the circuit runs, and therefore, if the device can execute the circuit in Fig.~\ref{fig:second_derivative} (a) while maintaining sufficient overall fidelity, it is advantageous to utilize Fig.~\ref{fig:second_derivative} (a) and Eq.~(\ref{eq:second_derivative_with_ancilla}).
Otherwise the low-depth version should be used to obtain meaningful result.

\subsection{Evaluation of $\frac{\partial}{\partial \theta_a}\frac{\partial}{\partial \theta_b}\frac{\partial E }{\partial \theta_c}$}
\begin{figure}
    \centering
    \includegraphics[width=0.9\linewidth]{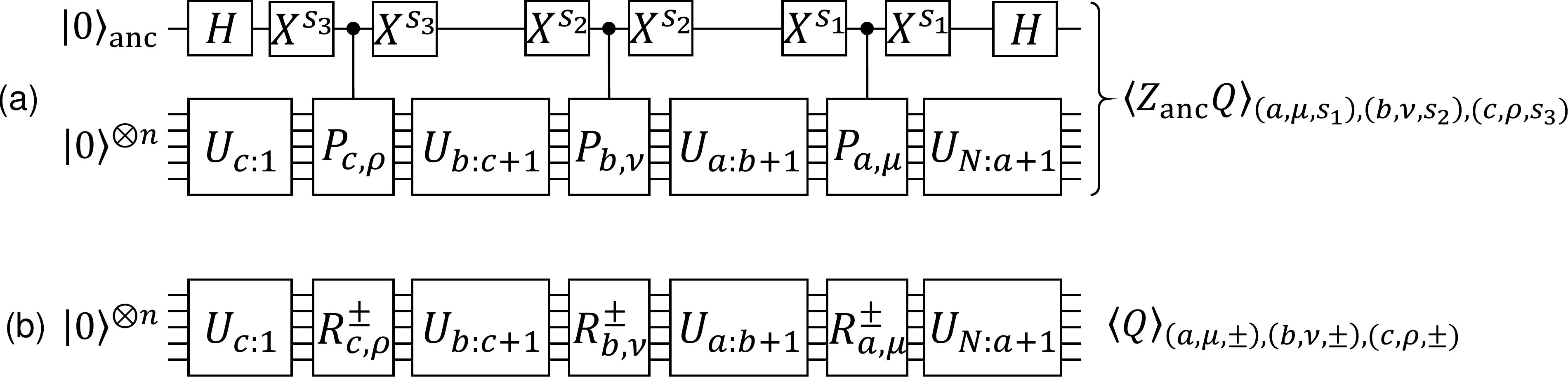}
    \caption{\label{fig:third_derivative} Quantum circuit to evaluate $\frac{\partial}{\partial \theta_a}\frac{\partial}{\partial \theta_b}\frac{\partial}{\partial \theta_c} E(\theta^*(x),x)$. For the definition of the notations, refer to Fig.~\ref{fig:second_derivative}. (a) Ancilla-based approach. See Eq.~(\ref{eq:third_derivative_ancilla}) for detailed procedure. (b) Low-depth version of (a). See Eq.~(\ref{eq:third_derivative_low-depth}) for detailed procedure.
    }
\end{figure}

Here, we show how to evaluate $\frac{\partial}{\partial \theta_a}\frac{\partial}{\partial \theta_b}\frac{\partial E}{\partial \theta_c} $. 
It should be clear how one can extend the method to the order higher than the third.
$\frac{\partial}{\partial \theta_a}\frac{\partial}{\partial \theta_b}\frac{\partial E}{\partial \theta_c}$ can be written down as,
\begin{align}
    &\frac{\partial}{\partial \theta_a}\frac{\partial}{\partial \theta_b}\frac{\partial E(\theta^*(x),x)}{\partial \theta_c}  \nonumber \\
    &= 2\sum_{\mu,\nu,\rho}\sum_{Q\in\mathcal{P}} h_Q(x) g_{a,\mu}g_{b,\nu}g_{c,\rho}\mathrm{Re}\left[\bra{\phi_{(a,\mu),(b,\nu),(c,\rho)}(\theta^*(x))}Q\ket{\psi(\theta^*(x))}\right. \nonumber \\
    &\qquad\qquad\qquad\qquad\qquad\qquad\qquad+
    \braket{\phi_{(a,\mu),(b,\nu)}(\theta^*(x))|Q|\phi_{(c,\rho)}(\theta^*(x))}
    \nonumber \\
    &\qquad\qquad\qquad\qquad\qquad\qquad\qquad+
    \braket{\phi_{(a,\mu),(c,\rho)}(\theta^*(x))|Q|\phi_{(b,\nu)}(\theta^*(x))}
    \nonumber \\
    &\qquad\qquad\qquad\qquad\qquad\qquad\qquad+
    \left.\braket{\phi_{(b,\nu),(c,\rho)}(\theta^*(x))|Q|\phi_{(a,\mu)}(\theta^*(x))}\right].\label{eq:nabla_theta_3_E}
\end{align}
Each term can be evaluated with the circuit in Fig.~\ref{fig:third_derivative} (a):
\begin{align}
    \mathrm{Re}\left[\bra{\phi_{(a,\mu),(b,\nu),(c,\rho)}(\theta^*(x))}Q\ket{\psi(\theta^*(x))}\right] &= \expect{Z_{\mathrm{anc}}Q}_{(a,\mu,0),(b,\nu,0),(c,\rho,0)},\nonumber \\
    \mathrm{Re}\left[\bra{\phi_{(a,\mu),(b,\nu)}(\theta^*(x))}Q\ket{\phi_{(c,\rho)}(\theta^*(x))}\right] &= \expect{Z_{\mathrm{anc}}Q}_{(a,\mu,0),(b,\nu,0),(c,\rho,1)},\nonumber \\
    \mathrm{Re}\left[\bra{\phi_{(a,\mu),(c,\rho)}(\theta^*(x))}Q\ket{\phi_{(b,\nu)}(\theta^*(x))}\right] &= \expect{Z_{\mathrm{anc}}Q}_{(a,\mu,0),(b,\nu,1),(c,\rho,0)},\nonumber \\
    \mathrm{Re}\left[\bra{\phi_{(b,\nu),(c,\rho)}(\theta^*(x))}Q\ket{\phi_{(a,\mu)}(\theta^*(x))}\right] &= \expect{Z_{\mathrm{anc}}Q}_{(a,\mu,1),(b,\nu,0),(c,\rho,0)}.\label{eq:third_derivative_ancilla}
\end{align}
The circuits can also be reduced to the low-depth version using the same strategy \cite{Mitarai2019a}.
The low-depth circuit for $\frac{\partial}{\partial \theta_a}\frac{\partial}{\partial \theta_b}\frac{\partial E}{\partial \theta_c}$ is shown in Fig.~\ref{fig:third_derivative} (b).
The circuit can evaluate each term of the summation by the following formula.
\begin{align}
    &-2\mathrm{Re}\left[
    \braket{\phi_{(a,\mu),(b,\nu),(c,\rho)}(\theta^*(x))|Q|\psi(\theta^*(x))}+
    \braket{\phi_{(a,\mu),(b,\nu)}(\theta^*(x))|Q|\phi_{(c,\rho)}(\theta^*(x))}\right.\nonumber\\
    &\left.\quad\quad\quad+\braket{\phi_{(a,\mu),(c,\rho)}(\theta^*(x))|Q|\phi_{(b,\nu)}(\theta^*(x))}
    +\braket{\phi_{(b,\nu),(c,\rho)}(\theta^*(x))|Q|\phi_{(a,\mu)}(\theta^*(x))}
    \right]\nonumber \\
    &= \expect{Q}_{(a,\mu,+),(b,\nu,+),(c,\rho,+)} - \expect{Q}_{(a,\mu,-),(b,\nu,-),(c,\rho,-)}\nonumber \\ 
    &\quad\quad\quad
    + \expect{Q}_{(a,\mu,-),(b,\nu,-),(c,\rho,+)} + \expect{Q}_{(a,\mu,-),(b,\nu,+),(c,\rho,-)}+\expect{Q}_{(a,\mu,+),(b,\nu,-),(c,\rho,-)}\nonumber\\
    &\quad\quad\quad
    -\expect{Q}_{(a,\mu,-),(b,\nu,+),(c,\rho,+)}-\expect{Q}_{(a,\mu,+),(b,\nu,-),(c,\rho,+)}-\expect{Q}_{(a,\mu,+),(b,\nu,+),(c,\rho,-)}
    \label{eq:third_derivative_low-depth}
\end{align}
Notice the pattern in the signs associating $\expect{Q}_{(a,\mu,\pm),(b,\nu,\pm),(c,\rho,\pm)}$, that is, the signs are determined in the parity of $\pm$ appearing in the subscript.
Higher order derivatives of $E$ with respect to $\theta$ can be evaluated by following the strategy described in this and the previous section.

Similar to the previous case, the low-depth version doubles the number of the circuit runs, and therefore, the same argument about the tradeoff between the noise and the number of the circuit runs also applies in this case.

\subsection{Evaluation of $\frac{\partial}{\partial \theta_a}\frac{\partial}{\partial \theta_b}\cdots\frac{\partial }{\partial \theta_c} \frac{\partial E}{\partial x_i}$}
\label{sec:theta_x_general_derivative}
$\frac{\partial}{\partial \theta_a}\frac{\partial}{\partial \theta_b}\cdots\frac{\partial }{\partial \theta_c} \frac{\partial E}{\partial x_i}$ can be measured with the same protocol as described in the previous sections; one can substitute $h_Q$ with $\frac{\partial h_Q}{\partial x_i}$.
With this substitution, Eqs.~(\ref{eq:nabla_theta_2_E}) and (\ref{eq:nabla_theta_3_E}) give us the analytical formula for $\frac{\partial}{\partial \theta_a}\frac{\partial}{\partial \theta_b}\cdots\frac{\partial }{\partial \theta_c} \frac{\partial E}{\partial x_i}$, where each term in the summation can be evaluated with the same procedure.
Also, $\frac{\partial}{\partial \theta_a}\frac{\partial}{\partial x_i}\frac{\partial E}{\partial x_k}$, which appears in Eq.~(\ref{eq:energy_third_derivative}), can be measured using the same strategy.

\section{Computational cost} \label{sec:cost}
Here, we give a comparison between the presented algorithm and the numerical differentiation to take $d$-th energy derivatives.
We specifically discuss the cost for each step in the overall algorithm presented in \ref{sec:overall} for quantum chemistry problem, where a problem Hamiltonian and its derivatives always has $O(n^4)$ terms as expressed by Eq. (\ref{eq:electron_hamiltonian}), as it is the main target of the proposed algorithm.
This analysis should be easy to be extended to more general cases.
We safely assume that classical computational cost involved in algorithms is much smaller than the quantum ones and thus can be ignored, since usual classical computation is much faster than sampling from the NISQ devices.

First, we ignore the cost of performing the VQE (Step 1), since both of the numerical differentiation and the presented algorithm need this step.
Step 2, 5, and 6 is merely classical, and therefore, we also ignore the cost for this part.
Step 3 of the presented algorithm requires us to run a quantum computer $O(n^4 /\epsilon^2)$ times to estimate each term up to an additional error of $\epsilon$.
For Step 4, it takes $O(n^4 \sum_{k=0}^{d-1} (N_\theta^{d-k})/\epsilon^2) = O(n^4 N_\theta^{d}/\epsilon^2)$ runs.
Note that even for the calculation of the terms involving $x$- and $\theta$-differentiation a the same time, such as $\frac{\partial }{\partial \theta_a}\frac{\partial E}{\partial x_i}$, we only need to perform measurement of the $\theta$-derivatives of each term, such as $\frac{\partial}{\partial \theta_a}\bra{\psi(\theta)}c_i^\dagger c_j^\dagger c_k c_l\ket{\psi(\theta)}$, in the Hamiltonian on the quantum device and then combine them by $\sum_{ijkl}\frac{\partial h_{ijkl}(x)}{\partial x}\frac{\partial}{\partial \theta_a}\bra{\psi(\theta)}c_i^\dagger c_j^\dagger c_k c_l\ket{\psi(\theta)}$ where $x$-differentiated $h(x)$ is computed classically. 
Therefore, $N_x$ does not appear in the number of NISQ runs.
Overall, the cost for the quantum part of the computation scales as $O(n^4 N_\theta^{d}/\epsilon^2)$.

Let us compare this cost against the numerical differentiation.
More specifically, we consider the case of evaluating the second derivative by the formula,
\begin{align}\label{eq:fd}
\frac{\partial^2 E^*(x)}{\partial x_i^2} \approx \frac{E^*(x+h) + E^*(x-h) - 2E^*(x)}{h^2},
\end{align}
where $h>0$.
Let $\tilde{E^*}(x)$ be the estimate of $E^*(x)$ obtained by measuring the state $\ket{\psi(\theta^*(x))}$.
If $|\tilde{E^*}(x) - E^*(x)|\leq \epsilon_E$, the precision of Eq.~(\ref{eq:fd}) is \cite{Hildebrand1987},
\begin{align}
\left|\frac{\partial^2 E^*(x)}{\partial x_i^2} - \frac{\tilde{E^*}(x+h) + \tilde{E^*}(x-h) - 2\tilde{E^*}(x)}{h^2}\right| \leq O(h^2) + O(\epsilon_E/h^2).
\end{align}
For classical computation where the source of $\epsilon_E$ is mainly the round-off error, the second term of the right-hand side is usually negligibly small for a decent $h$.
On the other hand, for the VQE, this term is the leading factor for the precision, since $\tilde{E^*}(x)$ has to be calculated by sampling.
To achieve $|\tilde{E^*}(x) - E^*(x)|\leq \epsilon_E$ with high probability, we need to run the quantum computer for $O(1/\epsilon_E^2)$ times.
Therefore, if we want to achieve the precision of $\epsilon$ for $\frac{\partial^2 E^*(x)}{\partial x_i^2}$, we need $O(1/(\epsilon^2 h^4))$ samples from $\ket{\psi(\theta^*(x))}$, and overall cost is $O(n^4 N_x^2/(\epsilon^2 h^4))$.
This scaling with respect to the error and the dimension of $x$ is clearly worse than that of the presented method, whose scaling is $O(1/\epsilon^2)$.

\section{Derivatives of excited state energy}
The generation of excited states can be a powerful application of the VQE, because the classical computation, despite the recent significant improvement in the theory and the computational power, still suffers in the calculation of them \cite{Lischka2018}.
Among the several proposals \cite{Endo2018b, Higgott2018, McClean2017b, Nakanishi2018} to generate excited states with the VQE, we adopt the one proposed in Refs.~\cite{Higgott2018, Endo2018b} to compute the derivatives of the excited state energy.

The algorithm \cite{Higgott2018,Endo2018b} works as follows.
First, we find the ground state of the given Hamiltonian $H_0(x)$, which we denote by $\ket{\psi^{(0)}(\theta^{(0)}(x))}$, where $\theta^{(0)}(x)$ is the optimal parameter for the ground state ($\theta^*(x)$ in the previous sections).
Then, we iteratively define a Hamiltonian $H_r(x)$ for $r=1,2,\cdots$ as,
\begin{align}
    H_r(x) := H_0(x) + \sum_{s=0}^{r-1}\beta_{s} \ket{\psi^{(s)}(\theta^{(s)}(x))}\bra{\psi^{(s)}(\theta^{(s)}(x))},
\end{align}
where $\ket{\psi^{(r)}(\theta^{(r)}(x))}$ is the ground state of $H_r(x)$.
If $\beta_s$ is sufficiently large, each $H_r(x)$ has $r$-th excited state of the original Hamiltonian, $H_0(x)$, as its ground state.
Therefore, by finding the ground state of each $H_r(x)$, one can generate the series of excited states of $H_0(x)$.
We assume $\ket{\psi^{(r)}(\theta)} = U^{(r)}(\theta)\ket{0}$, where $U^{(r)}(\theta)$ has the same structure as $U(\theta)$ in previous sections
In this algorithm, it is also assumed that we have a device which can measure the overlap between
$\ket{\psi^{(r)}(\theta)}$ and $\ket{\psi^{(s)}(\varphi)}$, that is, we assume we can measure $|\braket{\psi^{(r)}(\theta)|\psi^{(s)}(\varphi)}|^2$.
Let the expectation value of $H_r(x)$ with respect to the state $\ket{\psi^{(r)}(\theta)}$ be $E_r(\theta,x)$; $E_r(\theta,x) = \braket{\psi^{(r)}(\theta)|H_r(x)|\psi^{(r)}(\theta)}$.
We define the optimal energy by $E_r^*(x) = E_r(\theta^{(r)}(x),x)$.

The task is to compute the derivatives such as $\frac{\partial^2 E^*_r}{\partial x_i\partial x_j}$.
Since $E^*_r$ is the ground state energy for $H_r$, the formulae in the previous sections, Eqs.~(\ref{eq:energy_first_derivative}), (\ref{eq:energy_second_derivative}), and (\ref{eq:energy_third_derivative}) can be adapted for this task.
The only difference from that of the actual ground state is that the derivative of the Hamiltonian, such as $\frac{\partial^2 H_r}{\partial x_i\partial x_j}$, cannot be computed classically.
For example, the expression of the first derivative of the Hamiltonian is,
\begin{align}
    \frac{\partial H_r}{\partial x_i}(x) &= \frac{\partial H_0}{\partial x_i}(x) + \sum_{s=0}^{r-1} \beta_s \left(\frac{\partial \ket{\psi^{(s)}(\theta^{(s)}(x))}}{\partial x_i} \bra{\psi^{(s)}(\theta^{(s)}(x))} + \mathrm{h.c.}\right)\nonumber \\
    &= \frac{\partial H_0}{\partial x_i}(x) + \sum_{s=0}^{r-1} \sum_{a}\beta_s\frac{\partial \theta^{(s)}_a}{\partial x_i}(x)\left(\ket{\partial_a \psi^{(s)}(\theta^{(s)}(x))}\bra{\psi^{(s)}(\theta^{(s)}(x))} + \mathrm{h.c.}\right).\label{eq:Hr_first_derivative}
\end{align}
In the analytical expression for $\frac{\partial E^*_r}{\partial x_i}$ (Eq.~(\ref{eq:energy_first_derivative})), $\frac{\partial H_r}{\partial x_i}(x)$ appears as the expectation value with respect to $\ket{\psi^{(r)}(\theta^{(r)}(x))}$; $\braket{\psi^{(r)}(\theta^{(r)}(x))|\frac{\partial H_r}{\partial x_i}(x)|\psi^{(r)}(\theta^{(r)}(x))}$.
As for $\frac{\partial^2 E^*_r}{\partial x_i\partial x_j}$ (Eq.~(\ref{eq:energy_second_derivative})), it appears as $\mathrm{Re}\left[\braket{\psi^{(r)}(\theta^{(r)}(x))|\frac{\partial H_r}{\partial x_i}(x)|\partial_a\psi^{(r)}(\theta^{(r)}(x))}\right]$.
The quantities that cannot be computed classically in Eqs.~(\ref{eq:energy_first_derivative}) to (\ref{eq:energy_third_derivative}) are the terms which involves the inner product between the states, such as $\mathrm{Re}[\braket{\psi^{(r)}(\theta^{(r)}(x))|\partial_a\psi^{(s)}(\theta^{(s)}(x))}\braket{\psi^{(s)}(\theta^{(s)}(x))|\psi^{(r)}(\theta^{(r)}(x))}]$.
However, notice that if the condition  $\braket{\psi^{(r)}(\theta^{(r)}(x))|\psi^{(s)}(\theta^{(s)}(x))} = 0$ holds for all $x$, which we naturally expect at the optimal parameter, we obtain by differentiating with respect to $x$ the both hand side of the equation,
\begin{align}
\sum_{a}\frac{\partial \theta^{(s)}_a}{\partial x_i}(x)\mathrm{Re}[\braket{\psi^{(r)}(\theta^{(r)}(x))|\partial_a\psi^{(s)}(\theta^{(s)}(x))}\braket{\psi^{(s)}(\theta^{(s)}(x))|\psi^{(r)}(\theta^{(r)}(x))}] = 0.
\end{align}
We can therefore ignore the inner-product terms for the evaluation of derivatives of excited state energy and utilize the same procedure as the ground state energy in this case.
Likewise, the inner-product terms that appear in the higher-order derivatives can also be ignored when the orthogonality condition $\braket{\psi^{(r)}(\theta^{(r)}(x))|\psi^{(s)}(\theta^{(s)}(x))} = 0$ is satisfied.
The so-called subspace search VQE method~\cite{Nakanishi2018}, which guarantees the orthogonality condition of the resultant states, can be advantageous to obtain such condition.

\section{Numerical simulation}
We provide proof-of-principle numerical simulations using the electronic Hamiltonian of the hydrogen molecule.
The Hamiltonians are calculated with the open source library PySCF \cite{PYSCF} and OpenFermion \cite{Openfermion}.
The simulation of quantum circuits are performed with Qulacs \cite{Qulacs}.

\subsection{Approximation of the potential energy surface}
\begin{figure}
    \centering
    \includegraphics[width=0.5\linewidth]{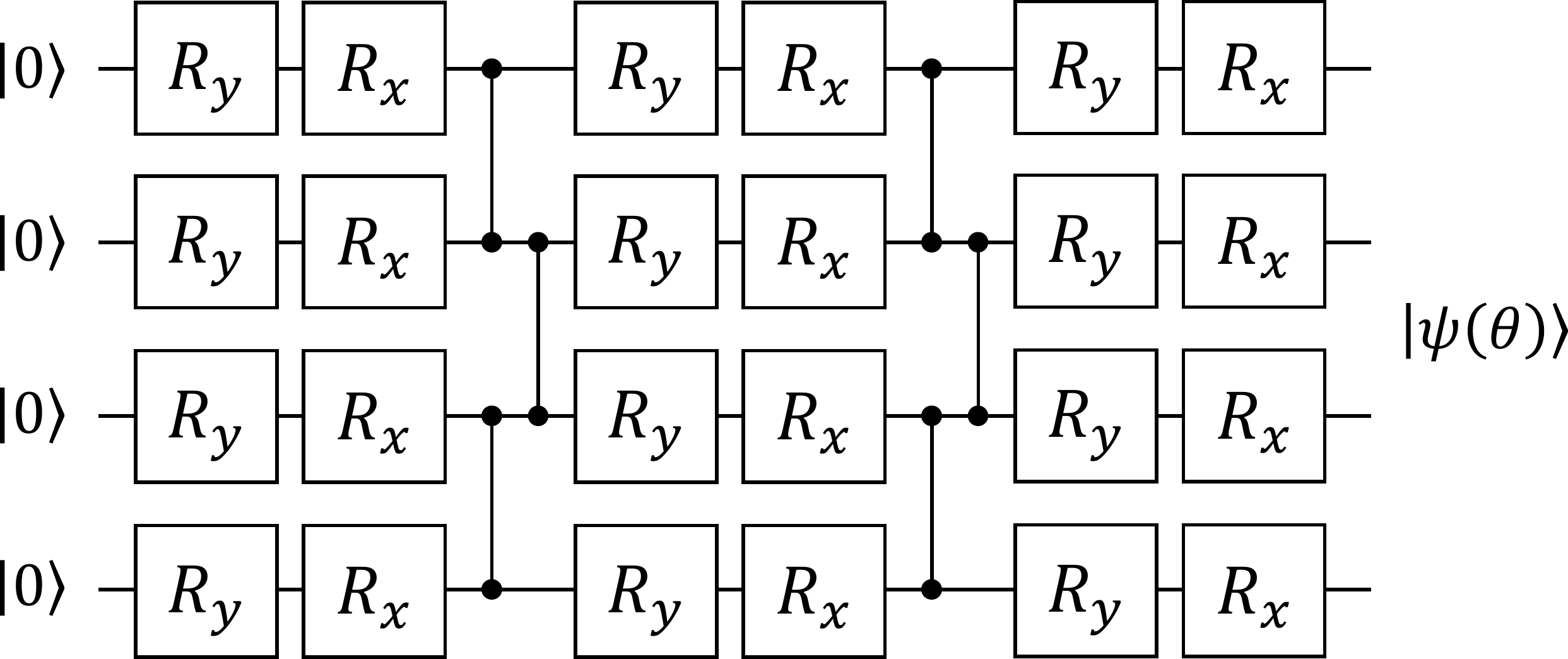}
    \caption{Ansatz used in simulations. $R_x$ and $R_y$ is single-qubit $x$ and $y$ rotation gate, respectively. The parameters $\theta$ are  implemented as rotation angles of $R_x$ and $R_y$.}
    \label{fig:ansatz}
\end{figure}

\begin{figure}
    \centering
    \includegraphics[width=0.45\linewidth]{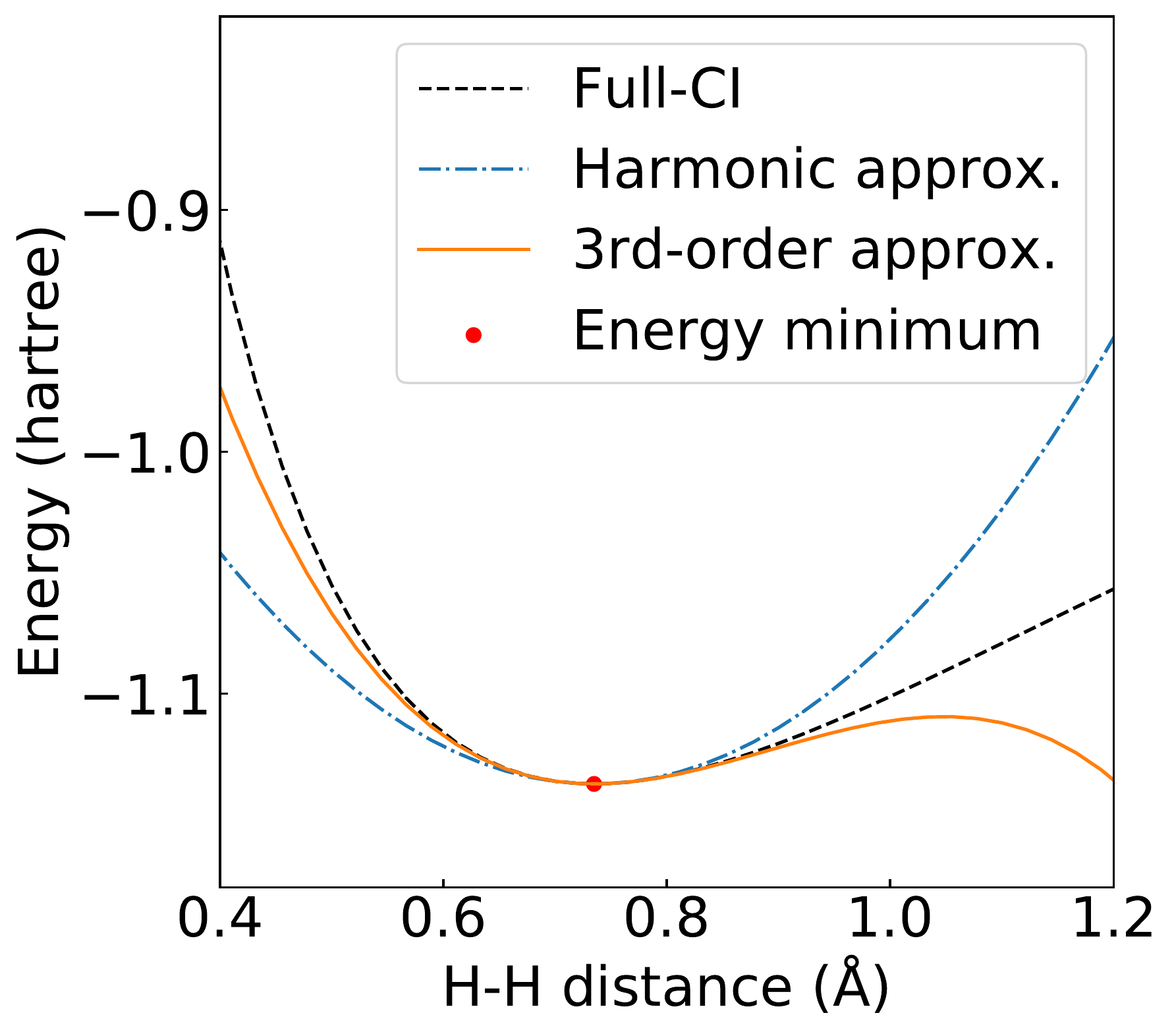}
    \caption{The harmonic and third-order approximation of the energy curve of the hydrogen molecule at the bonding distance, determined by the simulation of the VQE and the proposed method.}
    \label{fig:PES_approx}
\end{figure}

First, we perform a simulation of the VQE and the methods described in Sec.~\ref{sec:ground_state_derivative} with the Hamiltonian of a hydrogen molecule, calculated with the STO-3G basis set, at the bonding distance $r=0.735~\mathrm{\AA}$.
We use the ansatz circuit shown in Fig. \ref{fig:ansatz}, which is a variant of so-called hardware efficient circuit \cite{Kandala2017}.
The result of the simulation is shown as Fig. \ref{fig:PES_approx}.
The ansatz used in this simulation could achieve the exact ground state, which is called full configuration interaction (Full-CI) state in the context of chemistry, and therefore, we could draw the harmonic and the third-order approximation of the Full-CI energy as shown in Fig. \ref{fig:PES_approx}.
The harmonic approximation can be used to calculate the vibrational spectra of a molecule, and the third order approximation can be utilized for more accurate description of the vibration.

\subsection{Continuous determinination of the optimal parameter}
The response equation, Eq. (\ref{eq:theta_first_derivative}), can be used to determine the optimal paramter $\theta^*(x)$ from the one at the slightly different system parameter, $\theta^*(x+\delta x)$, that is, to the first order,
\begin{equation}\label{eq:euler_update}
    \theta^*(x+\delta x) \approx \theta^*(x) + \frac{\partial \theta^*}{\partial x}(x)\delta x,
\end{equation}
holds up to the additive error of $O(\delta x^2)$.
One can iteratively use the equation above, which resembles the Euler method, to obtain the optimal parameter $\theta_{\mathrm{Euler}}(x)\approx \theta^*(x)$ from some $\theta^*(x_0)$ in a range of $x$ around $x_0$ where the error term is sufficiently small.

\begin{figure}
    \centering
    \includegraphics[width=0.45\linewidth]{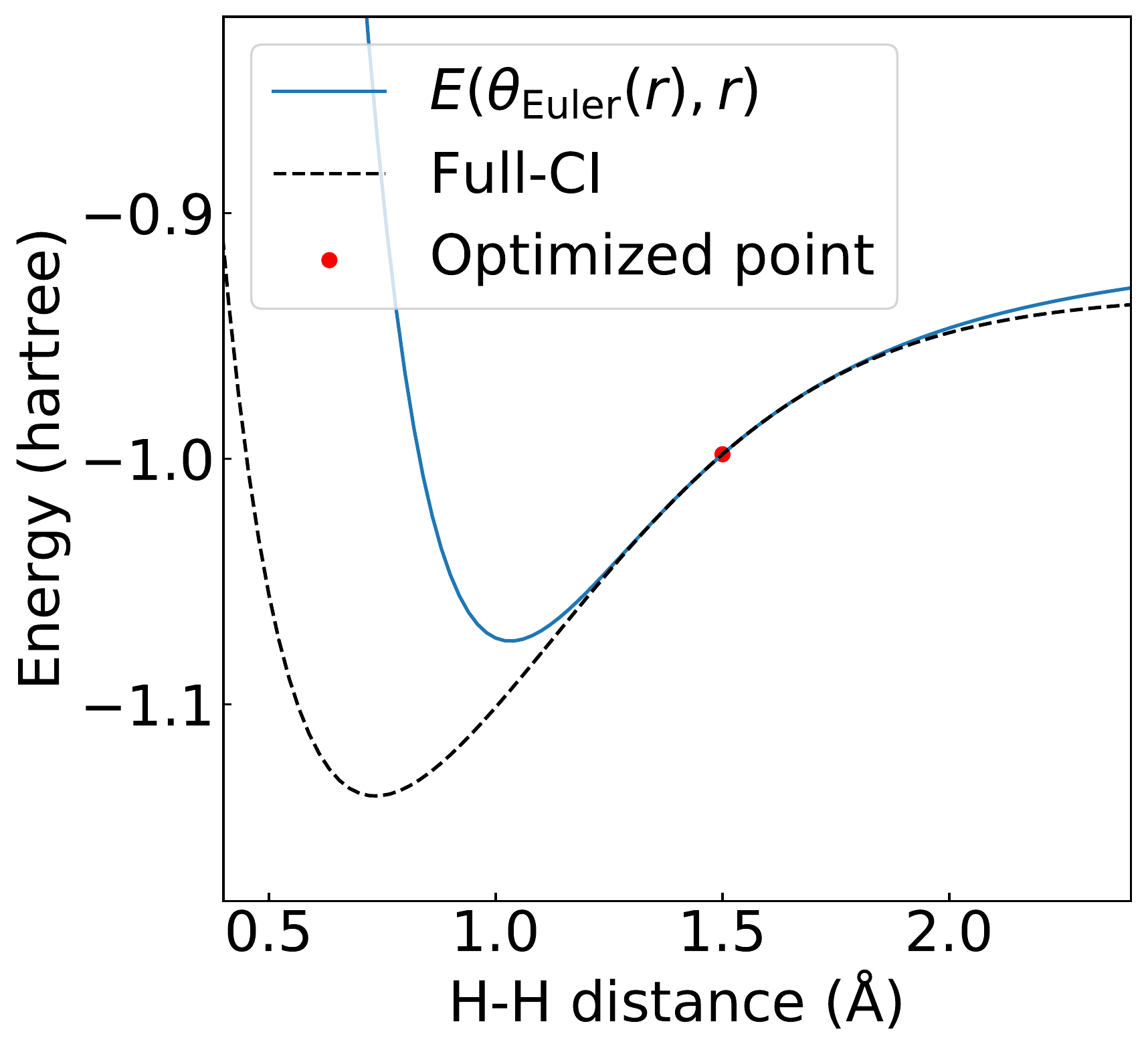}
    \caption{The evolution of the energy expectation value when the circuit parameters are evolved according to Eq. (\ref{eq:euler_update}), starting from the optimized parameter at interatomic distance of $1.5~\mathrm{\AA}$. }
    \label{fig:euler_update}
\end{figure}

We demonstrate this parameter update in Fig. \ref{fig:euler_update}, by plotting the energy expectation value of $\ket{\psi(\theta_{\mathrm{Euler}}(r))}$ where $\theta_{\mathrm{Euler}}(r)$ is determined with the iterative use of Eq. (\ref{eq:euler_update}), starting from the optimal parameter at $r=1.5~\mathrm{\AA}$ with the step size $\delta r = 0.02$ $\mathrm{\AA}$.
Eq. (\ref{eq:euler_update}) resembles the Euler method, therefore we refer to the update of $\theta$ according to Eq. (\ref{eq:euler_update}) as so in Fig. \ref{fig:euler_update}.
It can be seen that in the range where $r$ is sufficiently close to the optimized point, $r=1.5~\mathrm{\AA}$, the parameter determined from Eq. (\ref{eq:euler_update}) is almost optimal, but as $r$ goes far from the optimized point, they deviate fast from the optima.
This is because, as the error of the each update accumulates, the state becomes a non-variational state, that is, $\frac{\partial E}{\partial \theta} \neq 0$, which results in the breakdown of the response equation, Eq. (\ref{eq:theta_first_derivative}), where we assumed $\frac{\partial E}{\partial \theta} = 0$.
This method should be useful for drawing potential energy surfaces using the VQE, because it can reduce the time for the optimization of the circuit parameter by predicting the optimal parameter from the one determined with the slightly different system.
We belive it can also be combined with the parameter interpolation approach proposed in Ref. \cite{Mitarai2019}.

\section{Conclusion}
We have described a methodology for computing the derivatives of the ground state and the excited state energy.
We have shown the straight-forwardly constructed quantum circuit to measure the quantities necessary for the calculation of energy derivatives, and also the low-depth version of the circuit.
The low-depth protocol for the measurement makes them suitable for the use in NISQ devices.
We also performed numerical simulations of the proposed method, which validates the correctness.
This work enables to extract the physical properties of the quantum system under investigation, thus widening the application range of the NISQ devices.

\textit{Note added -} During the final revision of this article, we have become aware of recent article \cite{OBrien2019} which also describes a methodology for evaluating energy derivatives. Their work is based on perturbative sum-over-state approach and differs from our methods which provides a way to evaluate analytic derivatives. 

\begin{acknowledgments}
KM thanks for METI and IPA for its support through MITOU Target program.
KM is supported by JSPS KAKENHI No. 19J10978.
WM thanks for KAKENHI No. 18K14181.
This work is also supported by MEXT Q-LEAP.
\end{acknowledgments}

\appendix
\section{Derivative of the optimal parameter}
We derive the expression for the derivatives of the optimal parameter, such as $\frac{\partial \theta^*_a(x)}{\partial x_i}$, by Taylor expansion.
Assume that at $x=\alpha$, we know the optimal parameter $\theta^*(\alpha)$
We perform Taylor expansion of $\ket{\psi(\theta^*(x))}$ and $H(x)$ around $\alpha$.
For $H(x)$, we obtain,
\begin{align}
H(\alpha + x) = H(\alpha) + \sum_i \frac{\partial H(\alpha)}{\partial x_i} x_i + \frac{1}{2}\sum_i \frac{\partial}{\partial x_i}\frac{\partial H(\alpha)}{\partial x_j} x_ix_j + \cdots. \label{appeq:hamiltonian_taylor_exp}
\end{align}
For $\ket{\psi(\theta^*(x))}$, we can expand it as follows.
\begin{align*}
    &\ket{\psi(\theta^*(\alpha+x))} \\
    &= \ket{\psi\left(
        \theta^*(\alpha)
        + \sum_i \frac{\partial \theta^*(\alpha)}{\partial x_i}x_i
        + \frac{1}{2}\sum_{i,j} \frac{\partial}{\partial x_i}\frac{\partial \theta^*(\alpha)}{\partial x_j}x_i x_j
        + \cdots
    \right)}\\
    &= \ket{\psi\left(\theta^*(\alpha)\right)}\\
    &\quad + \sum_a \ket{\partial_a \psi\left(\theta^*(\alpha)\right)}\left[\sum_i \frac{\partial \theta^*_a(\alpha)}{\partial x_i}x_i
    + \frac{1}{2}\sum_{i,j} \frac{\partial}{\partial x_i}\frac{\partial \theta^*_a(\alpha)}{\partial x_j}x_i x_j
    + \cdots\right]\\
    &\quad + \frac{1}{2}\sum_{a,b} \ket{\partial_a \partial_b \psi\left(\theta^*(\alpha)\right)}\left[\sum_i \frac{\partial \theta^*_a(\alpha)}{\partial x_i}x_i
    + \frac{1}{2}\sum_{i,j} \frac{\partial}{\partial x_i}\frac{\partial \theta^*_a(\alpha)}{\partial x_j}x_i x_j
    + \cdots\right]
    \\
    &\quad\quad\quad\quad\quad\quad\quad\quad\quad\quad\quad\quad
    \left[\sum_i \frac{\partial \theta^*_b(\alpha)}{\partial x_i}x_i
    + \frac{1}{2}\sum_{i,j} \frac{\partial}{\partial x_i}\frac{\partial \theta^*_b(\alpha)}{\partial x_j}x_i x_j
    + \cdots\right]\\
    &\quad +\cdots.
\end{align*}
When grouped by the order of $x$,
\begin{align}
    &\ket{\psi(\theta^*(\alpha+x))} \nonumber\\
    &= \ket{\psi\left(\theta^*(\alpha)\right)}\nonumber\\
    &\quad + \sum_{a,i} \frac{\partial \theta^*_a(\alpha)}{\partial x_i}\ket{\partial_a \psi\left(\theta^*(\alpha)\right)}x_i\nonumber\\
    &\quad + \frac{1}{2}\sum_{i,j} 
    \left[\sum_{a} \frac{\partial}{\partial x_i}\frac{\partial \theta^*_a(\alpha)}{\partial x_j}\ket{\partial_a \psi\left(\theta^*(\alpha)\right)} + 
    \sum_{a,b}\frac{\partial \theta^*_a(\alpha)}{\partial x_i}\frac{\partial \theta^*_b(\alpha)}{\partial x_j}\ket{\partial_a \partial_b \psi\left(\theta^*(\alpha)\right)}\right]x_i x_j\nonumber
    \\
    &\quad +
    \frac{1}{6}\sum_{i,j,k}\left[
    \sum_a \frac{\partial}{\partial x_i}\frac{\partial}{\partial x_j}\frac{\partial \theta^*_a(\alpha)}{\partial x_k}\ket{\partial_a \psi\left(\theta^*(\alpha)\right)}
    + 3 \sum_{a,b} \frac{\partial}{\partial x_i}\frac{\partial\theta^*_a(\alpha)}{\partial x_j}\frac{\partial\theta^*_b(\alpha)}{\partial x_j}\ket{\partial_a\partial_b \psi\left(\theta^*(\alpha)\right)}\right.\nonumber\\
    &\quad\quad\quad\quad\quad\quad\quad\quad\quad\quad\quad\quad\quad\quad\quad\left.
    + \sum_{a,b,c} \frac{\partial\theta^*_a(\alpha)}{\partial x_j}\frac{\partial\theta^*_b(\alpha)}{\partial x_j}\frac{\partial\theta^*_c(\alpha)}{\partial x_k}\ket{\partial_a\partial_b\partial_c \psi\left(\theta^*(\alpha)\right)}
    \right]x_ix_jx_k\nonumber\\
    &\quad +\cdots. \label{appeq:taylor_expansion}
\end{align}
We can derive a similar expression for $\ket{\partial_a \psi(\theta^*(\alpha+x))}$.
Now, we use the condition of Eq.~(\ref{eq:grad_vanish}) to derive the expression of the derivatives of $\theta^*(\alpha)$. 
Plugging Eqs.~(\ref{appeq:hamiltonian_taylor_exp}) and (\ref{appeq:taylor_expansion}) into Eq.~(\ref{eq:grad_vanish}) and imposing the coefficient of each order in $x$ to be zero, we get the analytical expression for the derivatives of $\theta^*(\alpha)$.
In the following two subsections, we derive the expression for the first and second derivatives of $\theta^*(\alpha)$.

\subsection{First derivative}
For the first order in $x$, we get
\begin{align}
    &\mathrm{Re}\left[\sum_{a} \left(\braket{\partial_b \psi\left(\theta^*(\alpha)\right)|H(\alpha)|\partial_a \psi\left(\theta^*(\alpha)\right)} + \braket{\psi\left(\theta^*(\alpha)\right)|H(\alpha)|\partial_b \partial_a \psi\left(\theta^*(\alpha)\right)}\right) \frac{\partial \theta^*_a(\alpha)}{\partial x_i}\right. \nonumber\\
    &\left. \quad\quad\quad\quad\quad\quad\quad\quad\quad\quad\quad\quad\quad + \braket{\partial_b \psi\left(\theta^*(\alpha)\right)|\frac{\partial H(\alpha)}{\partial x_i}| \psi\left(\theta^*(\alpha)\right)}\right] = 0. \label{appeq:x_first_order}
\end{align}
Note that the first term of Eq.~(\ref{appeq:x_first_order}) includes the Hessian of $E(\theta,x)$, that is,
\begin{align}
    \frac{\partial}{\partial \theta_b}\frac{\partial E(\theta^*(\alpha),\alpha)}{\partial \theta_a}= 2\mathrm{Re}\left[\left(\braket{\partial_b \psi\left(\theta^*(\alpha)\right)|H(\alpha)|\partial_a \psi\left(\theta^*(\alpha)\right)} + \braket{\psi\left(\theta^*(\alpha)\right)|H(\alpha)|\partial_b \partial_a \psi\left(\theta^*(\alpha)\right)}\right)\right].
\end{align}
Also, notice that,
\begin{align}
    \frac{\partial}{\partial \theta_b}\frac{\partial E(\theta, x)}{\partial x_i}
    &=2\mathrm{Re}\left[\braket{\partial_b \psi\left(\theta\right)|\frac{\partial H}{\partial x_i}(x)| \psi\left(\theta\right)}\right].
\end{align}
We define $\nabla_\theta^d\frac{\partial E}{\partial x_i}$ likewise.
Finally, we obtain,
\begin{align}
    \sum_{a}\frac{\partial}{\partial \theta_b}\frac{\partial E(\theta^*(\alpha),\alpha)}{\partial \theta_a}\frac{\partial \theta^*_a(\alpha)}{\partial x_i} = -\frac{\partial}{\partial \theta_b}\frac{\partial E(\theta^*(x), x)}{\partial x_i}, \label{appeq:theta_first_derivative}
\end{align}
which is Eq.~(\ref{eq:theta_first_derivative}) of the main text.
Note that we expect the matrix $\frac{\partial}{\partial \theta_b}\frac{\partial E(\theta^*(\alpha),\alpha)}{\partial \theta_a}$ to be positive definite because $\theta^*(\alpha)$ is a local minimum, and thus Eq.~(\ref{appeq:theta_first_derivative}) is solvable to obtain $\frac{\partial \theta^*_a(\alpha)}{\partial x_i}$.

\subsection{Second derivative}
The second derivative, $\frac{\partial}{\partial x_i}\frac{\partial \theta_a}{\partial x_j}$, is derived from the second order in $x$ of Eq.~(\ref{eq:grad_vanish}).
We have, from Eq.~(\ref{eq:grad_vanish}), for all $i,j$ and $c$,
\begin{align}
    &\mathrm{Re}\left[\frac{1}{2}\sum_{a,b} \frac{\partial \theta^*_a(\alpha)}{\partial x_i}\frac{\partial \theta^*_b(\alpha)}{\partial x_j}(
    \braket{\partial_c \psi\left(\theta^*(\alpha)\right)|H(\alpha)|\partial_a\partial_b \psi\left(\theta^*(\alpha)\right)} + \braket{\partial_a\partial_b\partial_c \psi\left(\theta^*(\alpha)\right)|H(\alpha)|\psi\left(\theta^*(\alpha)\right)}\right.\nonumber\\
    &\quad\quad\quad\quad\quad\quad\quad\quad\quad\quad\quad
    + \braket{\partial_a\partial_c \psi\left(\theta^*(\alpha)\right)|H(\alpha)|\partial_b\psi\left(\theta^*(\alpha)\right)}+\braket{\partial_b\partial_c \psi\left(\theta^*(\alpha)\right)|H(\alpha)|\partial_a\psi\left(\theta^*(\alpha)\right)})
    \nonumber\\
    &\quad\quad
    + \frac{1}{2}\sum_{a} \frac{\partial}{\partial x_i}\frac{\partial \theta^*_a(\alpha)}{\partial x_j}(\braket{\partial_c \psi\left(\theta^*(\alpha)\right)|H(\alpha)|\partial_a \psi\left(\theta^*(\alpha)\right)} + \braket{\partial_a\partial_c \psi\left(\theta^*(\alpha)\right)|H(\alpha)| \psi\left(\theta^*(\alpha)\right)})
    \nonumber\\
    &\quad\quad
    + \sum_{a} \frac{\partial \theta^*_a(\alpha)}{\partial x_j}\left(\braket{\partial_c \psi\left(\theta^*(\alpha)\right)|\frac{\partial H(\alpha)}{\partial x_j}|\partial_a \psi\left(\theta^*(\alpha)\right)} + \braket{\partial_a\partial_c \psi\left(\theta^*(\alpha)\right)|\frac{\partial H(\alpha)}{\partial x_j}| \psi\left(\theta^*(\alpha)\right)}\right)
    \nonumber\\
    &\quad\quad\left.
    +\frac{1}{2}\braket{\partial_c \psi\left(\theta^*(\alpha)\right)|\frac{\partial}{\partial x_i}\frac{\partial H(\alpha)}{\partial x_j}|\psi\left(\theta^*(\alpha)\right)}
    \right] = 0.
\end{align}
$\frac{\partial}{\partial\theta_a}\frac{\partial}{\partial\theta_b}\frac{\partial E}{\partial\theta_c}$, $\frac{\partial}{\partial\theta_a}\frac{\partial E}{\partial\theta_b}$ can be used to greatly simplify the above equation, which gives us, 
\begin{align}
    &\frac{1}{4}\sum_{a,b} \frac{\partial \theta^*_a(\alpha)}{\partial x_i}\frac{\partial \theta^*_b (\alpha)}{\partial x_j} \frac{\partial}{\partial\theta_c}\frac{\partial}{\partial\theta_a}\frac{\partial E (\theta^*(\alpha), \alpha)}{\partial\theta_b}
    + \frac{1}{4}\sum_{a} \frac{\partial}{\partial x_i}\frac{\partial \theta^*_a (\alpha)}{\partial x_j}\frac{\partial}{\partial\theta_c}\frac{\partial E (\theta^*(\alpha),\alpha)}{\partial\theta_a}\nonumber \\
    &\quad\quad+ \frac{1}{2}\sum_{a} \frac{\partial \theta^*_a(\alpha)}{\partial x_i}\frac{\partial}{\partial\theta_c}\frac{\partial}{\partial\theta_a}\frac{\partial E (\theta^*(\alpha),\alpha)}{\partial x_j}
    +\frac{1}{2}\mathrm{Re}\left[\braket{\partial_c \psi\left(\theta^*(\alpha)\right)|\frac{\partial}{\partial x_i}\frac{\partial H(\alpha)}{\partial x_j}|\psi\left(\theta^*(\alpha)\right)}\right] = 0.
\end{align}
This is equivalent to Eq.~(\ref{eq:theta_second_derivative})

\section{Derivatives of the ground state energy}
\subsection{First derivative}
The first derivative of the energy is calculated as,
\begin{align*}
    \frac{\partial E^*}{\partial x_i}(x) &= \frac{\partial}{\partial x_i}\bra{\psi(\theta^*(x))}H(x)\ket{\psi(\theta^*(x))},\\
    &= 2\mathrm{Re}\left[\bra{\psi(\theta^*(x))}H(x)\frac{\partial \ket{\psi(\theta^*(x))}}{\partial x_i}\right] + \bra{\psi(\theta^*(x))}\frac{\partial H}{\partial x_i}(x)\ket{\psi(\theta^*(x))}, \\
    &= 2\sum_a \frac{\partial \theta^*_a(x)}{\partial x_i}\mathrm{Re}\left[\bra{\psi(\theta^*(x))}H(x)\ket{\partial_a \psi(\theta^*(x))}\right] + \bra{\psi(\theta^*(x))}\frac{\partial H}{\partial x_i}(x)\ket{\psi(\theta^*(x))},
\end{align*}
and first term of the above equation vanishes by Eq.~(\ref{eq:grad_vanish}) of the main text. Thus, we get,
\begin{align}
    \frac{\partial E^*}{\partial x_i}(x) = \braket{\psi\left(\theta^*(x),x\right)|\frac{\partial H}{\partial x_i}(x)| \psi\left(\theta^*(x),x\right)} \label{appeq:energy_first_derivative},
\end{align}
which is Eq.~(\ref{eq:energy_first_derivative}) of the main text

\subsection{Second derivative}
Here, we derive the expression for the second derivative.
\begin{align}
&\frac{\partial}{\partial x_i}\frac{\partial E^*}{ \partial x_j}(x) \nonumber \\
&= \frac{\partial}{\partial x_i} \braket{\psi(\theta^*(x)) | \frac{\partial H}{\partial x_j}(x) |\psi(\theta^*(x))} \nonumber \\
&=
2\mathrm{Re}\left[
    \bra{\psi(\theta^*(x))} \frac{\partial H}{\partial x_j}(x) \frac{\partial \ket{\psi(\theta^*(x))}}{\partial x_i} 
\right] +
\bra{\psi(\theta^*(x))}
    \frac{\partial }{\partial x_i}\frac{\partial H}{\partial x_j}(x) 
\ket{\psi(\theta^*(x))} \nonumber\\
&=
2\sum_a \mathrm{Re}\left[
    \bra{\psi(\theta^*(x)) } \frac{\partial H}{\partial x_j}(x) \ket{\partial_a \psi(\theta^*(x))}
\right] \frac{\partial \theta_a^*}{\partial x_i}(x)
+ 
\braket{\psi(\theta^*(x)) | \frac{\partial }{\partial x_i}\frac{\partial H}{\partial x_j}(x) |\psi(\theta^*(x))} \nonumber \\
&= \sum_a \frac{\partial \theta^*_a}{\partial x_i}(x)\frac{\partial}{\partial \theta_a}\frac{\partial E}{\partial x_j}(\theta^*(x),x) + \bra{\psi(\theta^*(x))} \frac{\partial }{\partial x_i}\frac{\partial H}{\partial x_j}(x) \ket{\psi(\theta^*(x))}.\label{appeq:energy_second_derivative}
\end{align}
This is Eq.~(\ref{eq:energy_second_derivative}) of the main text.

\subsection{Third derivative}
Third derivative can be calculated as follows.
\begin{align}
    &\frac{\partial}{\partial x_i}\frac{\partial}{\partial x_j}\frac{\partial E^*}{ \partial x_k}(x) \nonumber \\
    &= \frac{\partial}{\partial x_i}\left(\sum_a \frac{\partial \theta^*_a}{\partial x_j}(x)\frac{\partial}{\partial \theta_a}\frac{\partial E}{\partial x_k}(\theta^*(x),x) + \bra{\psi(\theta^*(x))} \frac{\partial }{\partial x_j}\frac{\partial H}{\partial x_k}(x) \ket{\psi(\theta^*(x))}\right)\nonumber \\
    &=\sum_a \frac{\partial}{\partial x_i}\frac{\partial \theta^*_a}{\partial x_j}(x)\frac{\partial}{\partial \theta_a}\frac{\partial E}{\partial x_k}(\theta^*(x),x) 
    + \sum_a \frac{\partial \theta^*_a}{\partial x_j}(x) \frac{\partial}{\partial x_i}\left(\frac{\partial}{\partial \theta_a}\frac{\partial E}{\partial x_k}(\theta^*(x),x)\right)\nonumber\\
    &\quad + 2\mathrm{Re}\left[\frac{\partial\bra{\psi(\theta^*(x))}}{\partial x_i} \frac{\partial }{\partial x_j}\frac{\partial H}{\partial x_k}(x) \ket{\psi(\theta^*(x))}\right]
    + \bra{\psi(\theta^*(x))}\frac{\partial }{\partial x_i}\frac{\partial }{\partial x_j}\frac{\partial H}{\partial x_k}(x) \ket{\psi(\theta^*(x))}\nonumber \\
    &=\sum_a \frac{\partial}{\partial x_i}\frac{\partial \theta^*_a}{\partial x_j}(x)\frac{\partial}{\partial \theta_a}\frac{\partial E}{\partial x_k}(\theta^*(x),x) 
    + \sum_{a,b} \frac{\partial \theta^*_a}{\partial x_j}(x) \frac{\partial\theta^*_b}{\partial x_i}(x)\left(\frac{\partial}{\partial \theta_b}\frac{\partial}{\partial \theta_a}\frac{\partial E}{\partial x_k}(\theta^*(x),x)\right)\nonumber\\
    &\quad + 2\sum_a \frac{\partial \theta^*_a}{\partial x_j}(x)\mathrm{Re}\left[\bra{\partial_a\psi(\theta^*(x))}\frac{\partial}{\partial x_i}\frac{\partial H}{\partial x_k}\ket{\psi(\theta^*(x))}\right] \nonumber\\
    &\quad + 2\sum_a\frac{\partial \theta^*_a}{\partial x_i}(x)\mathrm{Re}\left[\bra{\partial_a \psi(\theta^*(x))} \frac{\partial }{\partial x_j}\frac{\partial H}{\partial x_k}(x) \ket{\psi(\theta^*(x))}\right]
    + \bra{\psi(\theta^*(x))}\frac{\partial }{\partial x_i}\frac{\partial }{\partial x_j}\frac{\partial H}{\partial x_k}(x) \ket{\psi(\theta^*(x))}.\label{appeq:energy_third_derivative}
\end{align}
This expression (Eq.~(\ref{appeq:energy_third_derivative})) can be simplified so as to avoid the explicit calculation of $\frac{\partial}{\partial x_i}\frac{\partial \theta^*_a}{\partial x_j}$.
By multiplying $\frac{\partial \theta_a}{\partial x_k}$ to Eq.~(\ref{eq:theta_first_derivative}) and $\frac{\partial }{\partial x_i}\frac{\partial \theta_a}{\partial x_j}$ to Eq.~(\ref{eq:theta_second_derivative}), and combining them, we obtain,
\begin{align}
    &\frac{\partial}{\partial x_i}\frac{\partial \theta_a}{\partial x_j}\frac{\partial}{\partial \theta_a}\frac{\partial E}{\partial x_k}=\nonumber \\
    &\sum_{b,c}\frac{\partial}{\partial \theta_a}\frac{\partial}{\partial \theta_b}\frac{\partial E}{\partial \theta_c}\frac{\partial \theta_a}{\partial x_k}\frac{\partial \theta_b}{\partial x_i}\frac{\partial \theta_c}{\partial x_j} + 2\sum_b \frac{\partial}{\partial \theta_a}\frac{\partial}{\partial \theta_b} \frac{\partial E}{\partial x_j}\frac{\partial \theta_b}{\partial x_i}\frac{\partial \theta_a}{\partial x_k} + 2\mathrm{Re}\left[\braket{\partial_a \psi|\frac{\partial}{\partial x_i}\frac{\partial H}{\partial x_j}|\psi}\right]\frac{\partial \theta_a}{\partial x_k}.  
\end{align}
This yields,
\begin{align}
    &\frac{\partial}{\partial x_i}\frac{\partial}{\partial x_j}\frac{\partial E^*}{ \partial x_k}(x) \nonumber \\
    &=\sum_{a,b,c}\frac{\partial}{\partial \theta_a}\frac{\partial}{\partial \theta_b}\frac{\partial E}{\partial \theta_c}(\theta^*(x),x)\frac{\partial \theta_a}{\partial x_i}(x)\frac{\partial \theta_b}{\partial x_j}(x)\frac{\partial \theta_c}{\partial x_k}(x) + \bra{\psi(\theta^*(x))}\frac{\partial }{\partial x_i}\frac{\partial }{\partial x_j}\frac{\partial H}{\partial x_k}(x) \ket{\psi(\theta^*(x))} \nonumber \\
    &\quad + \sum_{a,b}\left[ \frac{\partial \theta^*_a}{\partial x_i}(x) \frac{\partial\theta^*_b}{\partial x_j}(x)\frac{\partial}{\partial \theta_b}\frac{\partial}{\partial \theta_a}\frac{\partial E}{\partial x_k}(\theta^*(x),x)
    +\frac{\partial \theta^*_a}{\partial x_k}(x) \frac{\partial\theta^*_b}{\partial x_i}(x)\frac{\partial}{\partial \theta_b}\frac{\partial}{\partial \theta_a}\frac{\partial E}{\partial x_j}(\theta^*(x),x)\right.\nonumber\\
    &\left.\qquad\qquad\qquad\qquad\qquad\qquad +\frac{\partial \theta^*_a}{\partial x_j}(x) \frac{\partial\theta^*_b}{\partial x_k}(x)\frac{\partial}{\partial \theta_b}\frac{\partial}{\partial \theta_a}\frac{\partial E}{\partial x_i}(\theta^*(x),x)
    \right]\nonumber\\
    &\quad + 2\sum_a \left[\frac{\partial \theta^*_a}{\partial x_i}(x)\mathrm{Re}\left[\bra{\partial_a \psi(\theta^*(x))} \frac{\partial }{\partial x_j}\frac{\partial H}{\partial x_k}(x) \ket{\psi(\theta^*(x))}\right] \right.\nonumber \\
    &\quad\quad\quad\quad\quad + \frac{\partial \theta^*_a}{\partial x_k}(x)\mathrm{Re}\left[\bra{\partial_a \psi(\theta^*(x))} \frac{\partial }{\partial x_i}\frac{\partial H}{\partial x_j}(x) \ket{\psi(\theta^*(x))}\right]\nonumber\\
    &\quad\quad\quad\quad\quad \left.+ \frac{\partial \theta^*_a}{\partial x_j}(x)\mathrm{Re}\left[\bra{\partial_a\psi(\theta^*(x))}\frac{\partial}{\partial x_k}\frac{\partial H}{\partial x_i}(x)\ket{\psi(\theta^*(x))}\right]\right],\label{appeq:energy_third_derivative_simplified}
\end{align}
which is Eq.(\ref{eq:energy_third_derivative}) of the main text.

\end{document}